\documentclass[aps, twocolumn, showpacs, preprintnumbers, superscriptaddress, amsmath, amssymb, prl]{revtex4-1}
\usepackage{amssymb}
\usepackage{mathrsfs}
\usepackage{amsmath}
\usepackage{graphicx}
\usepackage{bm}
\usepackage{epstopdf}
\usepackage{float}
\usepackage{multirow}
\usepackage{enumerate}
\usepackage{array}
\usepackage{color}
\usepackage[colorlinks=true, letterpaper=true, pdfstartview=FitV, linkcolor=blue, citecolor=blue, urlcolor=blue]{hyperref}

\makeatletter
\newcommand{\Rmnum}[1]{\expandafter\@slowromancap\romannumeral #1@}
\makeatother
\begin{document}
\title{Symmetry Dictionary on Charge and Spin Nonlinear Responses for All Magnetic Point Groups with Nontrivial Topological Nature}

\author{Zhi-Fan Zhang}
\affiliation{School of Physical Sciences, University of Chinese Academy of Sciences, Beijing 100049, China}

\author{Zhen-Gang Zhu}
\email{zgzhu@ucas.ac.cn}
\affiliation{School of Physical Sciences, University of Chinese Academy of Sciences, Beijing 100049, China}
\affiliation{School of Electronic, Electrical and Communication Engineering, University of Chinese Academy of Sciences, Beijing
100049, China}
\affiliation{CAS Center for Excellence in Topological Quantum Computation, University of Chinese Academy of Sciences, Beijing 100190,
China}

\author{Gang Su}
\email{gsu@ucas.ac.cn}
\affiliation{Kavli Institute for Theoretical Sciences, University of Chinese Academy of Sciences, Beijing 100190, China}
\affiliation{School of Physical Sciences, University of Chinese Academy of Sciences, Beijing 100049, China}
\affiliation{CAS Center for Excellence in Topological Quantum Computation, University of Chinese Academy of Sciences, Beijing 100190,
China}

\begin{abstract}
Recently, charge or spin nonlinear transport with nontrivial topological properties in crystal materials has attracted much
attention. In this paper, we perform a comprehensive symmetry analysis for all 122 magnetic point groups (MPGs) and provide a useful dictionary for charge and spin nonlinear transport from Berry curvature dipole, Berry connection polarization and Drude term with nontrivial topological nature. The results are obtained by making a full symmetry investigation on matrix representations of six nonlinear response tensors. We further identify every MPG that can accommodate two or three of the nonlinear tensors. The present work gives a solid theoretical basis for overall understanding the second-order nonlinear responses in realistic materials.
\end{abstract}

\maketitle

\emph{Introduction.}---
The general matrix representations of a response tensor is very important to discuss whether the corresponding physical phenomena exist or
not in the materials with given point group symmetries, including for instance, the linear Hall conductivity tensor in anomalous Hall effect
\cite{Grimmer1993, Seemann2015}, the magnetoelectric pseudo-tensor in magnetoelectric effect (or  Edelstein effect) \cite{He2020}.
In 2015,  Sodemann and Fu \cite{Sodemann2015} proposed a nonlinear Hall effect, being directly proportional to the dipole moment of
the Berry curvature over occupied states, which is defined as a rank-two pseudo-tensor, the Berry curvature dipole (BCD), that is allowed in 18 classical point groups.
This effect is in fact the second-order charge nonlinear Hall effect contributed by the BCD.
Many materials such as bilayer or few-layer WTe$_2 $ \cite{Du2018,Qiong2019,Kang2019}, half-Heusler
alloy CuMnSb \cite{Shao2020}, ferroelectric-like metal LiOsO$_{3}$ \cite{Xiao2020} and so on\cite{Zhang_2022,Battilomo2019,He2021,Chen2019} are found to accommodate this effect.
Subsequently,  Oiwa and Kusunose \cite{Oiwa2022} proposed a systematic analysis for identifying essential parameters in various linear
and nonlinear response tensors to decompose the response tensors into the model-independent and -dependent parts using the Keldysh
formalism  and the Chebyshev polynomial expansion method \cite{Joao2019}.
Another intrinsic contribution to the second-order charge nonlinear Hall effect has been proposed, which can be described by
a rank-three tensor $ \chi^{\text{int}}$ relating  to the Berry connection polarization (BCP) \cite{Gao2014}. 
Recently, Liu \textit{et al.} \cite{Liu2021} showed the constraints of common point group operations on the in-plane tensor elements of
the BCP,  only including $ \chi_{xyy} $  and $ \chi_{yxx} $.
The list of magnetic point groups (MPGs) classified by the existence or nonexistence of second-order response tensor including
BCD, BCP  and Drude effect is shown directly in Table I in Ref. \cite{Wang2021}.
For the second-order spin transport, there were the symmetric analysis under time reversal symmetry (TRS) or space inversion (IS)
\cite{Hamamoto2017,Zhang2021}, revealing that 16 magnetic point groups satisfying $ \mathcal{PT} $ symmetry can make second-order spin
conductivity tensor non-zero \cite{Hayami2022} using the multipole classification \cite{Yatsushiro2021}.

 All second-order charge and spin response tensors can be defined as rank-three tenors that are characterized by three $3\times 3$ matrices. To the best of
our knowledge, the theories mentioned above do not describe the second-order charge and spin effects in a compact form as a whole, 
nor matrix representations of the general tensors subjected to various MPGs.
In this work, we present full matrix representations of the second-order charge and spin response tensors for all 122 MPGs. The powerful
role of this symmetry analysis rests in the fact that the general second-order response effects, such as the BCD, BCP and Drude can be
determined directly from their MPGs whether they exist, coexist or do not. More importantly, we identify, for the first time, a contribution of BCP in
addition to BCD and Drude for the second-order spin current. On all accounts the symmetry analyses are necessary for an overall understanding of topological nonlinear transport as well as searching candidate materials for hosting nonlinear responses of charge and spin.

\begin{figure}[tb]
	\centering
	\includegraphics[width=1\linewidth]{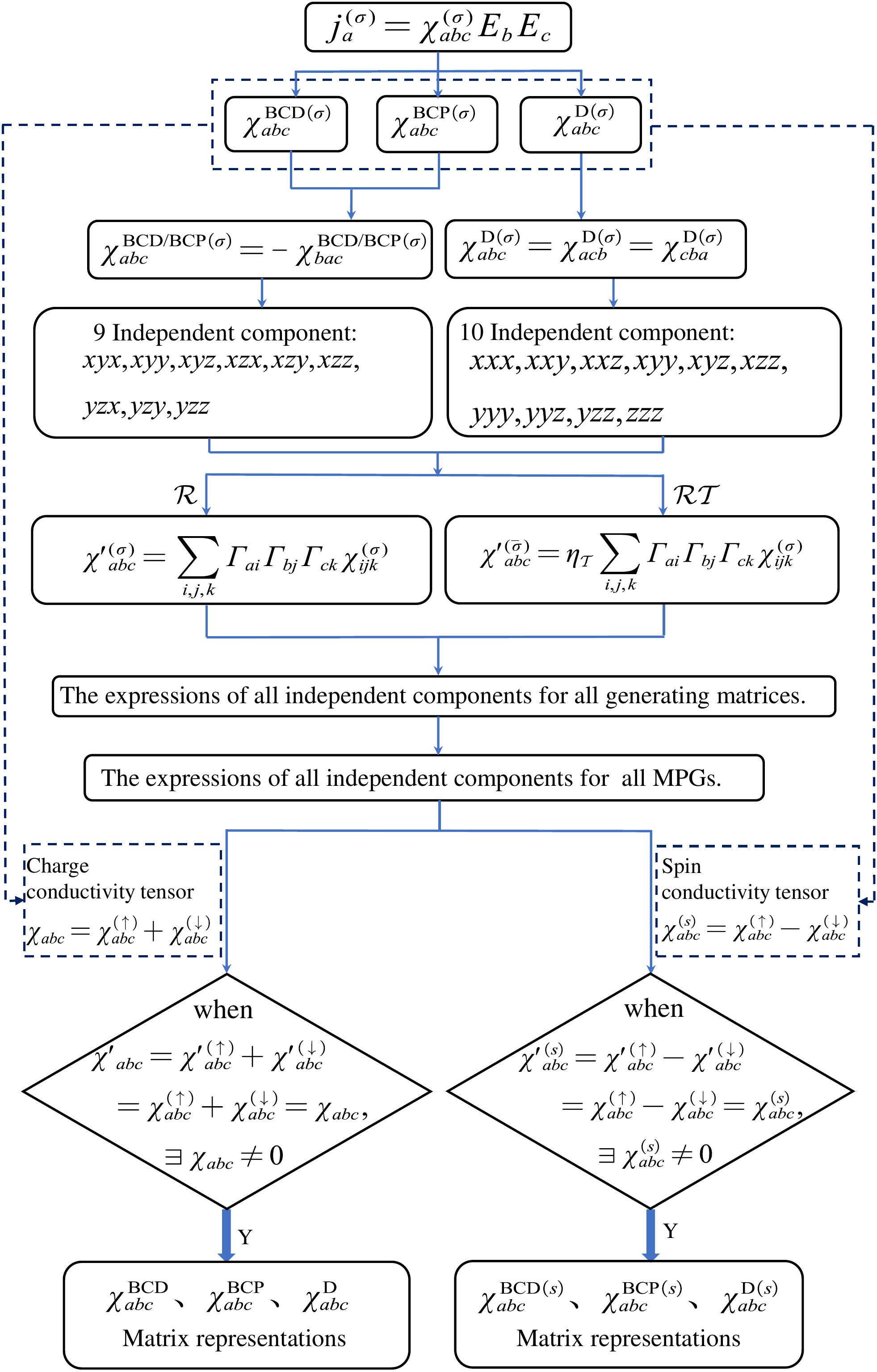}
	\caption{The calculation flow chart of matrix representations  of second-order charge and spin response tensors.}
	\label{fig1}
\end{figure}

\emph{The second-order charge and spin response tensors.}---
Under the driving electric field, we can obtain the second-order response tensor $\chi_{abc}^{(\sigma)}$ with the help of
semi-classical Boltzmann transport theory under the relaxation-time approximation \cite{Mahan2000,Abrikosov1988}, which can be
expressed as
	\begin{align}
		j_a^{(\sigma)}&=\chi _{abc}^{(\sigma)}E_bE_c,	
		\label{eq1}\\
	\chi _{abc}^{(\sigma)}&=\chi _{abc}^{\text{BCD}(\sigma)}+\chi _{abc}^{\text{BCP}(\sigma)}+\chi _{abc}^{\text{D}(\sigma)},
	\label{eq2}
\end{align}
where $a,b,c \in \{x,y,z\}$, $\sigma$ = $\uparrow$ or $\downarrow$ is spin index of electron and Eq. (\ref{eq1}) implies
the sum over the indices $ b $ and $ c $. Eq. (\ref{eq2}) includes three different contributions, i.e., the BCD $\chi
_{abc}^{\text{BCD}(\sigma)}$ \cite{Sodemann2015}, BCP $\chi _{abc}^{\text{BCP}(\sigma)}$ \cite{Wang2021,Liu2021} and Drude
$\chi _{abc}^{\text{D}(\sigma)}$ \cite{Gao2019,Zelezny2021,Itahashi2022,Lesne2022} terms for the second-order response. In Supplementary Materials \cite{SM},
we give a detailed derivation and explicit expressions of these terms.
$\chi _{abc}^{\text{BCP}(\sigma)}$ and $\chi _{abc}^{\text{BCD}(\sigma)}$  both have 9 independent components because of the $ a
\leftrightarrow b $  anti-symmetries \cite{Wang2021,Du2021}, while for the Drude term $\chi _{abc}^{\text{D}(\sigma)}$,
as any two indices are interchangeably symmetric, it has 10 independent components (shown in Fig. \ref{fig1}).
 Then the second-order charge current and the spin current can be expressed as $ j_a =j_a^{(\uparrow)}+j_a^{(\downarrow)}$
 and $ j_a^{(s)}=(\hbar/2e)(j_a^{(\uparrow)}-j_a^{(\downarrow)}) $, respectively \cite{Kane2005}. Notably, the second-order charge response tensor
 can be written as $\chi_{abc}=\chi_{abc}^{\uparrow}+\chi_{abc}^{\downarrow}$, which has BCD, BCP and Drude terms, whereas the spin response tensor can be defined as $\chi_{abc}^{(s)}=\chi_{abc}^{\uparrow}-\chi_{abc}^{\downarrow}$. For convenience, we may call these terms as the
 spin-dependent BCD, BCP  and Drude ones ($ \chi^{\text{BCD}(s)}  $, $ \chi^{\text{BCP}(s)}  $ and $ \chi^{\text{D}(s)}  $).

\emph{Calculation process.}---
In order to better describe magnetic materials, we introduce the magnetic point groups (also called Shubnikov groups) $\mathcal{M}$,
which can be divided into three classes (see Refs. \cite{Shubnikov1964,Newnham2005,Birss1963,Batanouny2008}):
\begin{enumerate}[I.]
	\renewcommand{\labelenumi}{Class \theenumi.}
	\setlength{\itemindent}{3em}
	\item $\mathcal{M}$ does not contain the time reversal operation $\mathcal T$;
	\item $\mathcal{M}$ contains $\mathcal{T}$ as an element;
	\item $\mathcal{M}$ contains $\mathcal{T}$ only in combination with another symmetry element $ \mathcal{R} $, such as
$\mathcal{C}_2^z\mathcal{T}$, where the $ z $ means the axis of the rotation is along the $ z $ direction in Cartesian coordinates.
\end{enumerate}

In three dimensions, there are  122 (32+32+58) MPGs, which are relisted in Supplementary Materials \cite{SM}. In order to facilitate
sorting, all subsequent MPG symbols adopt international symbols \cite{Batanouny2008}. The symbols of Class I MPGs do not contain the
prime, all the symbols of Class II are suffixed by $ 1' $, and the Class III contains letters or numbers (except $ 1 $) plus prime.  It
is then apparent that the Class I MPGs are equivalent to 32 classical point groups $ \mathcal{G}  $ ($\mathcal{R} \in \mathcal{G}
$), which can be used to describe nonmagnetic crystals. Or it corresponds to the crystals in some magnetic states, where each
atom has a certain magnetic moment, neither $\mathcal{T}$ nor $ \mathcal{RT}$ are symmetric operations. Paramagnetic or diamagnetic
materials with TRS can be described by the Class II. Class III must correspond to magnetic crystals, such as ferromagnetic,
antiferromagnetic or ferrimagnetic materials.
For all symmetric operators in 122 MPGs, we can use $ 3 \times 3 $ matrices to express them.  Any set of symmetry matrices from which
all symmetry matrices of a particular MPG may be obtained by multiplication, are known as a set of generating matrices $ \varGamma $
\cite{Birss1963,SM}, and similarly, the corresponding symmetry operators are known as the generators of the particular MPG. Then there are 18 generating matrices for all MPGs, which includes $ \varGamma^{n} $ and $ \underline{\varGamma}^n = \varGamma^n \mathcal{T}, n \in \{1,2,\cdots,9\}$ \cite{SM}.

For the second-order response tensors, the constraints from group symmetries on $\chi _{abc}^{(\sigma)} $ can be derived from
\begin{align}
	\chi _{abc}^{(\sigma')} =\eta_{\mathcal{T}} \sum_{i,j,k} \varGamma _{ai}^n \varGamma _{bj}^n \varGamma _{ck}^n\chi
_{ijk}^{(\sigma)},
	\label{eq3}
\end{align}	
where $\varGamma_{pq}^n$ ($ p=a,b,c; q=i,j,k  $) represent the element in row $ p $ and column $ q $ of the generating matrix  $
\varGamma^{n} $. The spin index $ \sigma'=\sigma$ and $ \eta_{\mathcal{T}} =1 $  are only
for Class I MPGs ($\mathcal{T}$ is not contained). 
While in Class II and III MPGs ($ \mathcal{T}$ is contained), spin is reversed, i.e. $\sigma'=\bar{\sigma}$. And 
the BCD term is even, $ \chi_{abc}^{\text{BCD}(\sigma)}(\mathbf{k}) =\chi_{abc}^{\text{BCD}(\sigma)}(\mathbf{-k})$, we have  $
\eta_{\mathcal{T}} = 1$; while the BCP and Drude terms are both odd with $ \eta_{\mathcal{T}} = - 1$.  
Thus, we can derive the
expressions with independent components for all MPGs in terms of the generating matrices (refer to \cite{SM}).
%
%
As the components of charge and spin response tensors are unchanged before and after the symmetry operation, we can find every nonzero component of the matrices so that we can get matrix representations of the corresponding tensors, as shown in Fig. \ref{fig1}.


\begin{figure}[tb]
	\centering
	\includegraphics[width=1\linewidth]{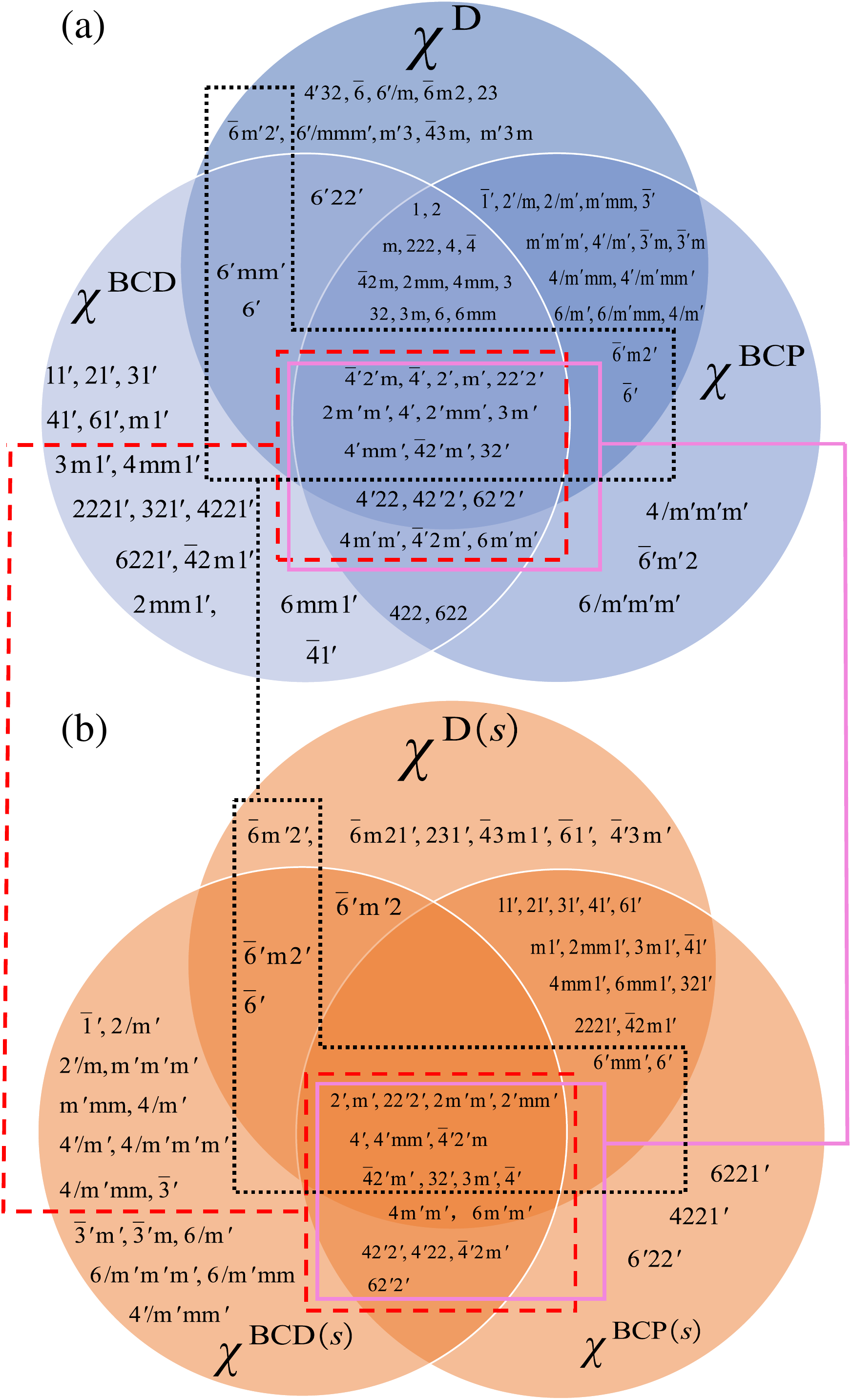}
	\caption{The classification and connection between the MPGs allowing for (a) charge and (b) spin conductivity tensors. The
red dash box connected by a dash line represents the MPGs where the $ \chi^{\text{BCD}} $ and the $ \chi^{\text{BCD}(s)} $ both exist,
while the pink solid and the black dot boxes represent the MPGs where the BCP and Drude terms coexist. }
	\label{fig2}
\end{figure}

\emph{Symmetry Analysis.}---
Fig. \ref{fig2}(a) shows all MPGs that allow for nonzero $\chi_{abc}^{\text{BCD}/\text{BCP}/\text{D}}$.
There are 16 Class I MPGs that make $\chi_{abc}^{\text{BCD}}$ non-zero, which has been discussed in Ref. \cite{Sodemann2015}.  It should be
noted that the group 23 ($T$)  and 432 ($ O $) restrict BCD to have only three equal diagonal components (similar to the
magnetoelectric pseudo-tensor in Ref. \cite{He2020}), whose corresponding second-order charge conductivity $ \chi_{abc} $ has to be zero due
to the $ a \leftrightarrow b $  anti-symmetry of the Levi-Civita symbol \cite{SM}.
Since the BCD term satisfies TRS, 16 Class II MPGs are found to accommodate the $ \chi^{\text{BCD}} $. In addition, there are
other 21 Class III MPGs that encompass non-zero BCD term. Because the BCP term is $\mathcal{T}$-odd, Class II cannot host nonzero BCP term, whereas 16
Class I and 37 Class III MPGs can give nonzero BCP terms.
As the Drude term $ \chi^{\text{D}} $ is symmetric when interchanging any two indices, the number of MPGs (allowing nonzero $ \chi^{\text{D}}$)
is a bit larger, namely, there are 18 Class I and 40 Class  III MPGs. All these results are presented in Table S3-S5 in Supplementary
Materials \cite{SM}.

According to whether the three terms $\chi^{\text{BCD}} $, $ \chi^{\text{BCP}} $ and $ \chi^{\text{D}} $ (which are represented by
three circles arranged counterclockwise) exist, coexist or not, we can divide the allowed MPGs into seven categories, as
illustrated by different overlapping areas of three circles in Fig. \ref{fig2}(a).
For instance, the lightest part of the lower left corner does not overlap the other two circles, implying that only $ \chi^{\text{BCD}}
$ nonzero  in the 16 Class II MPGs of $ 11', 21'$, etc., because the BCD term is $ \mathcal{T} $-even but the BCP and Drude terms are $
\mathcal{T} $-odd.
By contrast, there are $ 4/\text{m}'\text{m}'\text{m}' $, $ \bar{6}'\text{m}'2 $, $ 6/\text{m}'\text{m}'\text{m}' $ make $
\chi^{\text{BCP}} $ nonzero alone, and $ \bar{6}$, $ 6'/\text{m} $, etc. uniquely make  $ \chi^{\text{D}} $ nonzero.
Furthermore, the overlapping region of any two circles represents that two of three terms can coexist but the rest one is zero in some
special MPGs.
%
For example, some MPGs in the Class I and III   
accommodate the coexistence of $\chi^{\text{BCD}} $ and $ \chi^{\text{BCP}} $ (such as MPGs
$422, 4\text{m}'\text{m}'$), which is partly due to the fact that the two terms are nonzero in some combined operations (Ref. \cite{Liu2021}), e.g., $ \mathcal{C}_2^x \mathcal{T}$ and $ \mathcal{S}_4^x \mathcal{T} $.
Fortunately, we can distinguish them by use of different dependence on relaxation time, i.e. $ \chi^{\text{BCD}} \propto  \tau^2$ 
\cite{Sodemann2015}, and   $ \chi^{\text{BCP}} \propto  \tau^0 $ \cite{Wang2021,Liu2021}. 
The BCP and Drude terms are both nonzero only in 16 Class III MPGs, whereas the Drude term is proportional to $ \tau $ in this situation. It is
interesting to note that only three MPGs, $ 6'22',6'\text{m}\text{m}',6' $, satisfy the condition that the BCD and Drude terms coexist. Of particularly interesting is the central part of Fig. \ref{fig2}(a) where there are 29 MPGs to host the coexistence of all three terms. Thus we have an entire atlas of the second-order charge nonlinear responses under different MPGs.

Since the spin degree of freedom plays essential role in most materials, let us discuss the second-order spin response tensors for Class II
and Class III MPGs, and the results are given in Fig. \ref{fig2}(b).
It turns out that all Class II MPGs forbid the spin-dependent BCD term because $\chi^{\text{BCD}(\uparrow/\downarrow)} $ becomes
$\chi^{\text{BCD}(\downarrow/\uparrow)} $ in TRS, resulting in vanishing $\chi^{\text{BCD}(s)} $.
%
Among 58 Class III MPGs, 37 out of them accommodate the spin-dependent BCD contribution to the second-order spin current. There are
16 Class II and 21 Class III MPGs for $\chi^{\text{BCP}(s)} $ to be nonzero; while 18 Class II and 21 Class III for nonzero
$\chi^{\text{D}(s)} $.  
We would like to mention here that the spin-dependent BCP term is \textit{for the first time} proposed, which is also one of our central results.

We identify three categories in which the spin-dependent BCD, BCP and Drude terms exist uniquely (no coexistence), 
and there are 16, 3, and 6 MPGs,  respectively. 
%
Moreover, there are 4 MPGs ($42'2',4'22,\bar{4}'2\text{m}'$ and $62'2'$) allowing coexistence of the spin-dependent BCD and BCP.  
%
In the $11', 21'$, etc., the second-order spin currents are contributed both by the spin-dependent BCP and Drude contributions; 
while only three MPGs, $ \bar{6}'\text{m}'2 $, $ \bar{6}'\text{m}2' $  and $ \bar{6}'$, allow for the coexistence of the spin-dependent BCD and Drude terms when BCP has no
contribution. 
Similarly, the central part of Fig. \ref{fig2}(b) presents that the three terms are all nonzero in 14 Class III MPGs.
All these allowed Class II and Class III MPGs are shown in Table S6-S8 in Supplementary Materials \cite{SM}.

Importantly, we show in Fig. \ref{fig2} whether the second-order charge and spin currents can appear simultaneously or not. 
The red-dashed boxes indicate for the MPGs that belong to Class III, $ \chi^{\text{BCD}} $ and $ \chi^{\text{BCD}(s)}$ both
exist. 
For example, for the MPG $ \text{m}' $, the generating matrices are $ \varGamma^0 $ and $ \underline{\varGamma}^5 $ (=$\varGamma^5\mathcal{T} $). 
The former is a unitary matrix and the latter represents the operator $\mathcal P\mathcal{C}_2^z \mathcal T $, which makes the charge and spin
response tensors are all nonzero.
In addition, the MPGs outside the boxes do not allow coexistence of the BCD and spin-dependent BCD, but only one of them,
such as the MPG  $6'$ which allows $\chi^{\text{BCD}} $ but forbids $\chi^{\text{BCD}(s)} $  owing to the combination of generating matrices $ \varGamma^6$($ =\mathcal{C}_3^{1z}  $) and $\underline{\varGamma}^3$($ =\varGamma^3 \mathcal{T} = \mathcal{C}_2^z \mathcal{T}$). 
The MPGs containing $ \mathcal{PT} $ (for example $ \bar1', 2/\text{m}'$) only have nonzero $ \chi^{\text{BCD}(s)}$ (but zero $
\chi^{\text{BCP}(s)}$ and $ \chi^{\text{D}(s)}$), which reproduces the results in Ref. \cite{Hayami2022}. And the nonzero $
\chi^{\text{BCD}(s)}$ may be tested in many collinear magnets, such as LiFePO$_4$ \cite{ToftPetersen2015} and Cr$_2 $O$ _3 $
\cite{Brown2002}.
The pink-solid boxes encircle 18 MPGs that allow for the coexistence of second-order charge and spin currents from the BCP term.
Outside the pink-solid boxes are incompatible MPGs. The spin-dependent BCP term (zero charge BCP) can exist in 19 MPGs including 16 Class II and 3 Class III MPGs. 
In particular, three Class-III MPGs ($ 6'22'$, $6'\text{m}\text{m}'$ and $6'$) may be interesting since they host magnetic materials that may have wide spin-applications. 
The black-dotted boxes show the nonzero Drude term in second-order charge and spin responses. 
%
And outside the black-dotted boxes we can find four Class-II ($ \bar{6}\text{m}21', 231',\bar{4}3\text{m}1' $ and $ \bar{6}1' $) and one
Class-III ($ \bar{4}'3\text{m}' $) MPGs  in which only $\chi^{\text{D}(s)}$ is nonzero but the others ($\chi^{\text{BCD/BCP/D}}$ and$\chi^{\text{BCD(s)/BCP(s)}}$) are all zero. 
We would like to remark that these five MPGs are of particular interests as the materials with these MPGs can be ideal platforms to study the spin transport 
without involving charge transport. 


\emph{Matrix representation.}---
Besides the symmetry analysis, it is still necessary to obtain the specific matrix representations of the second-order charge  and spin
response tensors, $ \chi_{abc} $ and $ \chi_{abc}^{(s)} $, for all allowed MPGs (see Supplementary Materials \cite{SM}).
Here we give two examples in Table \ref{table1}.
Let us first study the MPG $ \text{m}1' $ (the $ 1'$  indicates that the system satisfies TRS) in bilayer or few-layer WTe$ _2 $ \cite{Brown1966}. For the BCD term in the second-order charge response tensor, the nonzero independent components are $xyx,xyy,xzz,yzz$. In two
or quasi-two dimensions, the driving electric field and the response current are both in the $ x-y $ plane, $ \text{m} $ (i.e., the
C$_{1v}$ group) is the only one MPG that has nonzero components (see \cite{SM}), and the maximum symmetry in which a second-order
current exists is the single mirror line under the TRS  \cite{Sodemann2015,footnote1}. In this case the second-order response charge
current contributed by the BCD is a transverse effect, being perpendicular to the direction of the external electric field.
In addition, many MPGs contain the combination of spatial and time reversal symmetry ($ \mathcal{PT} $), which forbids the BCD
contribution but allows for the BCP and Drude terms, such as the MPG $ 2'/m $ in antiferromagnetic tetragonal CuMnAs \cite{Wang2021} as another example.
According to Table \ref{table1}, the BCP term only has nonzero independent elements $ xyx, xyy,xzz,yzz $; while the Drude term
has the $ xxx,xxy,xyy,xzz,yyy,yzz$.
Intriguingly, this suggests the second-order charge current contributed by the BCP is a pure transverse effect, but the Drude term can
manifest itself as transverse and longitudinal components.
The matrix representations for the second-order charge (Table S9-S14) and spin  (Table S15-S19) response
tensors of all MPGs can be found in Supplementary materials \cite{SM}.

\begin{table}[h]
		\renewcommand\arraystretch{2}
		\caption{The nonzero independent components of charge response tensor for MPGs $ \text{m}1' $ and $  2'/\text{m}$. }
		\begin{tabular*}{8.5cm}{@{\extracolsep{\fill}}ccccc}
				\hline\hline
				MPG & $ \chi_{abc}^{\text{BCD}} $  & $ \chi_{abc}^{\text{BCP}} $  & $ \chi_{abc}^{\text{D}} $& Materials\\
				\hline
			\multirow{2}*{ $ \text{m}1' $ }& $xyx,xyy,xzz$ & \multirow{2}*{ zero}& \multirow{2}*{ zero}& \multirow{2}*{WTe$ _2 $}\\
			&$ yzz $\\
				\hline
  	        \multirow{2}*{ $ 2'/\text{m} $ }&  \multirow{2}*{zero}&$ xyx,xyy,xzz,$&$ xxx,xxy,xyy$ & \multirow{2}*{CuMnAs} \\
			& & $ yzz $ &$  xzz,yyy,yzz$\\
				\hline\hline
			\end{tabular*}
		\label{table1}
\end{table}

\emph{More materials.}---In addition to the materials mentioned above, we can inspect the database \cite{mpg1,mpg2} or existing
predicted structures \cite{Watanabe2018,Xu2020,Calugaru2021} to find corresponding materials for every MPG hosting the
second-order nonlinear charge and spin responses. On the other hand, we can also perform numerical calculations through known $ k\cdot p$  models \cite{Jiang2021,Tang2021} or first-principles calculations to find more materials with giant second-order effects. In 
Supplementary Materials \cite{SM} we provide some candidate materials for certain MPGs. 

\emph{Summary.}---
In this work, we show that both the second-order charge and spin response currents have three contributions, i.e., the BCD, BCP and Drude
terms, all of which can be written as rank-three tensors. Besides, a new term, i.e. the spin-dependent BCP tensor in second-order  spin response, is proposed in this work.  
To discuss the second-order nonlinear charge and spin responses, a detailed symmetry analysis and the matrix representations  of
six nonlinear response tensors are made for 122 MPGs. 
It turns out that we present a symmetry dictionary to find whether the second-order charge and spin response tensors can exist, coexist or not for a given MPG, and a few candidate magnetic materials are also suggested to detect the specific second-order nonlinear responses. 
Here we should point out that although this work does not include the scattering or electron-phonon interaction effects, some recent developments \cite{Watanabe2020,Michishita2022,Watanabe2021,Wang2022,Lahiri2022} make good complementary contributions to this intriguing topic, which together with this present work provide a comprehensive understanding on the second-order nonlinear charge and spin responses in realistic materials.

\emph{Acknowledgments.}---
The authors thank Dr. Da-Kun Zhou for helpful suggestions.
This work is supported in part by the NSFC (Grants No. 11974348 and No. 11834014), and the National Key R\&D Program of China (Grant No.
2018YFA0305800). It is also supported by the Fundamental Research Funds for the Central Universities, and the Strategic Priority
Research Program of CAS (Grants No. XDB28000000, and No. XDB33000000). Z.G.Z. is supported in part by the Training Program of Major Research plan of
the National Natural Science Foundation of China (Grant No. 92165105).


\begin{thebibliography}{51}
\bibitem [{foo()}]{footnote1}%
	\BibitemOpen
	\href@noop {} {\bibinfo  {journal} {In two dimensions, the components of
			charge response tensor are not zero for the MPG m, which only has a mirror
			symmetric operator. So the maximum symmetry that exists for a second-order
			charge current is a single mirror line}\ }\BibitemShut {NoStop}%
\end{thebibliography}
%

\end{document}